\documentclass[amsmath,amssymb,aps,floatfix,prl,superscriptaddress,preprint]{revtex4-1}

\usepackage[utf8]{inputenc} 
\usepackage[T1]{fontenc}    
\usepackage{amsfonts}       
\usepackage{booktabs}       
\usepackage{hyperref}       
\usepackage{url}            
\usepackage{microtype}      
\usepackage{nicefrac}       
\usepackage{paralist}       

\usepackage[dvipsnames]{xcolor}
\usepackage{graphicx}
\usepackage{float}
\usepackage{wrapfig}
\usepackage{ulem}

\usepackage{tikz}
\usetikzlibrary{arrows,shapes,angles,quotes}
\usetikzlibrary{decorations.pathreplacing}
\usepackage{fullpage}

\usepackage{lmodern}
\usepackage{ulem}

\begin{document}
	
	\title{On the Excess Entropy Scaling Law: a Potential Energy Landscape View}
	
	\author{Anthony Saliou}
	\affiliation{Université Grenoble Alpes, CNRS, Grenoble INP, SIMaP\\
		F-38000 Grenoble, France}
	\author{Philippe Jarry}
	\affiliation{C-TEC, Parc Economique Centr'alp, 725 rue Aristide Bergès, CS10027, Voreppe 38341 cedex, France }
	\author{Noel Jakse}
	\affiliation{Université Grenoble Alpes, CNRS, Grenoble INP, SIMaP\\
		F-38000 Grenoble, France}
	
	\begin{abstract}
		{
			The relationship between excess entropy and diffusion is revisited by means of large-scale computer simulation combined to supervised learning approach to determine the excess entropy for the Lennard-Jones potential. Results reveal that it finds its roots in the properties of the potential energy landscape (PEL). In particular the exponential law holding in the liquid is seen to be correlated with the landscape-influenced regime of the PEL while the fluid-like power-law corresponds to the free diffusion regime. 
		}
	\end{abstract}
	
	\maketitle
	
Understanding the link between dynamic properties of fluids and their underlying structure and thermodynamics \cite{Chapman1939,Eyring1939,Hansen2006} remains one of the major open issues in condensed matter physics. The link is formally embodied at the microscopic level in the Van Hove function \cite{VanHove1954} which led to the developments of the Mode-Coupling Theories \cite{Gotze1992} in the context of undercooled liquids \cite{Das2004}. It also yielded various generalisations of the Chapman-Enskog theory for hard-sphere fluids based on the time evolution of distribution functions \cite{Dyer2007} through integro-differential equations \cite{Martynov1992}. The role of thermodynamic quantities like entropy in describing the relaxation time or viscosity was shown in the Adam-Gibbs theory \cite{Adam1965} in glass-forming liquids, however Rosenfeld \cite{Rosenfeld1977,Rosenfeld1999}  was the first to reveal that the excess entropy,  $S_{ex}$, with respect to the ideal gas exhibits an exponential scaling law with diffusion, viscosity as well as thermal diffusion. More recent developments follow a line guided by the search of an underlying theoretical basis \cite{Dyre2018} of such a relation in the framework of isomorph theory with hidden scale invariance \cite{Dyre2015,Dyre2016} in which the entropy plays the central role \cite{Yoon2019}.  
	
The excess entropy scaling law was shown to be more general than expected initially: it applies for a wide range of systems and thermodynamic states (see Ref. \cite{Dyre2018}), and for instance for pure metals and alloys \cite{Hoyt2000} yielding a simple link between partial entropies and corresponding diffusions of the constitutive elements \cite{Pasturel2016}. It counts however a number of counter-examples \cite{Dyre2018} showing that it is more a semi-quantitative approach. It nevertheless possesses a quasi-universal character for systems having an interaction potential energy model $u(r)$  between particles of mass $m$ obeying Euler's law of corresponding states $u(r,\epsilon,\sigma) = \epsilon f(r/\sigma)$, where  $\epsilon$ and $\sigma$ are appropriate energy and length scales, respectively \cite{Rosenfeld1999}. Justifications are based \cite{Rosenfeld1977} on the ability of fluids with temperature $T$ and density $\rho$ to be generally well represented by a hard-sphere (HS) reference system \cite{Barker1976} of diameter $\sigma_{HS}$ that was shown to be valid beyond the class of Eulerian potentials \cite{Jakse2016}. It was further argued by Dzugutov \cite{Dzugutov1996} that with an appropriate scaling of the diffusion coefficient $D$ using the collision frequency within the Enskog theory for HS, $\Gamma=4\pi^2 g(\sigma_{HS}) \rho\sqrt{\pi k_B T/m}$, $k_B$ being Boltzmann's constant and $g(r)$ the pair-correlation function, leads to an exponential law, $exp(S_{ex})$, interpreted as being proportional to accessible microscopic states that leads to diffusion. In Rosenfeld's approach the scaling factor of $D$, \textit{i.e.} $\rho^{1/3}\sqrt{m/k_B T}$, is more general and does not depend on the underlying HS reference, while in Dzugutov's one, under the additional assumption that $S_{ex}\simeq S_2$, the two-body term of the excess entropy \cite{Baranyai1989}, the scaling law makes it possible to determine the diffusion solely from $g(r)$ accessible not only from simulation but also experimentally.  
	
The non-universal character of excess entropy law lies on the fact that entropy is basically a macroscopic equilibrium thermodynamic quantity while diffusion emerges from the rate of transition between underlying accessible microscopic states \cite{Dyre2018}. Nevertheless, a unified picture that may conciliates both these aspects is provided by the Potential-Energy Landscape (PEL) concept \cite{stillinger1995,sastry1998} in which the thermodynamics comes from the relative depths of the PEL minima while the dynamics is guided by their connectivity \cite{Wales2003}. The effect of complexity of the PEL on Rosenfeld's exponential law was investigated through a Gaussian random interacting potential and showed that it still holds for moderate disorder \cite{Seki2015}. A linear relationship between local  characteristics of the PEL and diffusivity assuming the excess entropy scaling was shown \cite{Chakraborty2006}. Nevertheless, a more general link between the PEL and excess scaling law is still missing to date. 

The aim of the present work is to provide a deeper insight on the excess entropy scaling law through the PEL concepts highlighted here for the Lennard-Jones (LJ) potential. The choice of the LJ model was guided by the fact that it leads to a quasi-universal character of the excess entropy scaling \cite{Rosenfeld1999}. This led us firstly to determine accurately the entropy over the largest fluid domain of the $\rho^*-T^*$ phase diagram, where $\rho^*=\rho\sigma^3$ and $T^*=k_BT/\epsilon$ and $\sigma$  being respectively the LJ length and energy scales. This non-trivial task  \cite{Jakse2016} is achieved by setting up a supervised Machine Learning (ML) approach using Artificial Neural Networks (ANN) \cite{Mehta2019,Schmidt2019} in order to combine databases covering different thermodynamic domains irrespective of the mathematical formulation of their equation of states (EOS). It was proven recently to be efficient to investigate the crystal nucleation by Monte-Carlo simulation taking entropy as a reaction coordinate \cite{Desgranges2018}. The second part consists in conducting large scale molecular dynamics (MD) simulations \cite{Smi2002} to generate an accurate, consistent and homogeneous dataset of self-diffusion coefficients spanning over the same range of the fluid phase diagram. Our findings show that the general characteristics of the PEL might be linked to the excess entropy scaling, \textit{i.e.} Rosenfeld's exponential law is seen to be correlated with the landscape influenced part of the PEL \cite{sastry1998} while the fluid-like power-law corresponds to its free diffusion regime. 

The entropy is determined in two steps from the basic thermodynamic relation $S_{ex}= (U_{ex}-A_{ex})/T$, where $U_{ex}$ is the excess internal energy and $A_{ex}$ the excess Helmholtz free energy. For the latter, many EOS were proposed in the literature \cite{Thol2016,Bell2019} that are valid over various domains of the phase diagram. Here, the Johnson, Zollweg and Gubbins EOS \cite{Johnson1993} for temperature domain $0.5 < T^* \leq 6.0$ and the Thol \textit{et al.} one for higher temperatures in the range $6.0 < T^*\leq 9.0$, was used. Considering densities from $0.005$ to $\rho^*=1.2$ allows us to cover a range of pressures up to $P^*=P\sigma^3\epsilon=50$. In this domain, using the respective EOS equations the dataset is composed of $16448$ values of $A^*_{ex}=A_{ex}/\epsilon$ for regularly spaced couples $(\rho^*,T^*)$ in the domain defined above. The ANN was built from densely connected Multi-Layer Perceptron (MLP) with a hyperbolic tangent activation function in a Feed-Forward (FF) configuration \cite{Hastie2009}. it was coded using Keras module from the TensorFlow Python package \cite{TensorFlow} in the regression mode. The supervised training is carried out using the macroscopic two component descriptor $(\rho^*;T^*)$ as the input, and the corresponding $A^*_{ex}$ as the output. For this, a randomized training set containing 80\% of the dataset was used, the remaining part being left for the test set. Optimization of the ANN configuration and hyperparameters used can be found in the Supplementary Material file \cite{Supplement}. Once trained, the ANN was used to predict $A^*_{ex}$ whatever the temperature and density inside the above defined domain allowing us to predict the excess entropy  with a high accuracy by simply using the internal energy from the corresponding MD simulation\cite{Supplement}.       

\begin{figure}[tb!]
	\centering
	\includegraphics[scale=0.7]{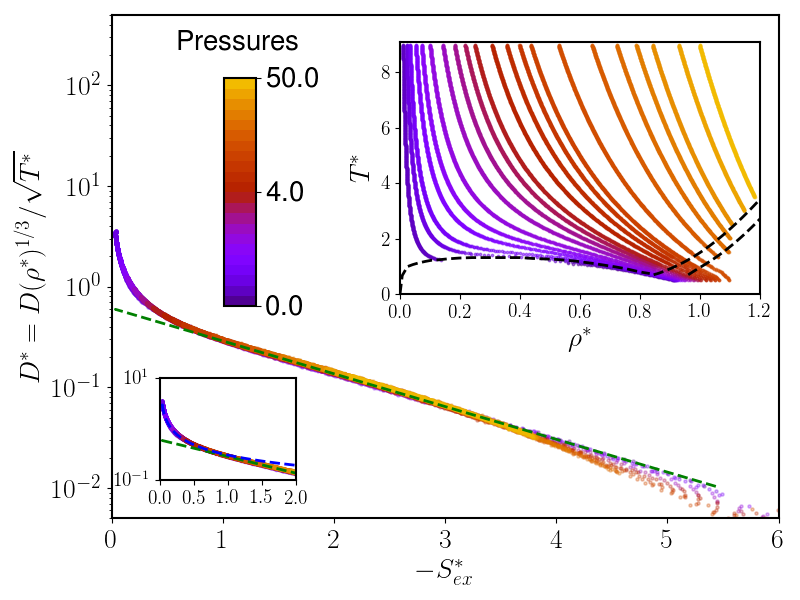}
	\caption{Scaled diffusion $D^*$ as a function of excess entropy $-S^*_{ex}=-S_{ex}/Nk_B$ along $22$ isobars shown in the upper right Inset. The green dashed line corresponds to a fit of the diffusion data in the range $1 \leq S^*_{ex}  \leq 3$ with Rosenfeld' law $0.61\exp(-0.751S^*_{ex})$ (see text). The considered pressures are $P^*=0$, $0.1$, $0.2$, $0.3$, $0.5$, $0.7$, $1$, $1.5$, $2$, $2.5$, $3$, $4$, $5$, $6$ , $7$, $10$, $15$, $20$, $25$, $30$, $40$, and $50$. The black dashed lines correspond to liquid-gas coexistence after Ref. \cite{Panagiotopoulos1994} below $\rho^*=0.8$ and the solid-liquid coexistence after Ref. \cite{Mastny2007} above. The lower left Inset shows the scaled diffusion in the dilute fluid range with small $-S^*_{ex}$. The blue dashed line corresponds to Rosenfeld's \cite{Rosenfeld1999} power law $0.305(-S_{ex})^{-2/3}$.}
	\label{fig:Fig1}
\end{figure}

The MD simulations were performed with $N=10768$ LJ particles in the isobaric-isothermal ensemble (NPT) within the Nosé-Hoover scheme \cite{Smi2002} using the LAMMPS code \cite{LAMMPS}. The truncated LJ was used with cut-off radius as large as $r_C = 4\sigma$ and standard long-range corrections \cite{Smi2002} were applied to pressures and energies. Equations of motions were integrated with velocity-verlet algorithm and a timestep $\delta t^*=0.001$. The chosen large value of $N$ reduces strongly the finite size effects so that corrections on the diffusion coefficients were not necessary \cite{Yeh2004}. For all the thermodynamic states, the system was equilibrated during $10^5$ time steps followed by a production run of $2\times10^5$ steps during which averaged properties were calculated. 

Fig. \ref{fig:Fig1} shows the evolution the reduced diffusion by means of Rosenfeld's scaling as a function of $-S^*_{ex}=-S_{ex}/Nk_B$ for all the simulated state points. The latter were obtained along $22$ isobars with a temperature steps of $0.02$ that cover quite homogeneously the considered phase diagram as, shown in the inset. In doing so, $-S^*_{ex}$  spans almost over the complete range considered for most of the pressures. Diffusion curves for all isobars collapse nicely onto a master curve for $-S^*_{ex}$ going from $0$ in the dilute gas phase to the dense liquid phase even up to roughly $3.9$ corresponding to the melting line \cite{Mastny2007} as will be seen below. Such a behavior is well-known and corresponds to Rosenfeld's findings commented several times and nicely reviewed very recently by Dyre \cite{Dyre2018}. The data shown in the curves have unprecedented low dispersion \cite{Bell2019}, which can be attributed to the fact that (i) the diffusion coefficients and corresponding potential energies are extracted from the same body of $17500$ large scale MD simulations, and (ii) the entropy is obtained consistently using a machine learning approach on accurate EOSs \cite{Johnson1993,Thol2016}. Following Rosenfeld \cite{Rosenfeld1999}, above melting, the excess entropy scaling shows a clear crossover around $-S^*_{ex}\simeq 1$ with a power law for dilute gas below $1$, \textit{i.e.} $0.305\times(-S_{ex})^{-2/3}$ and the exponential law $0.61*\exp(-0.751S^*_{ex})$ above, as shown in the lower left Inset of Fig. \ref{fig:Fig1}. The exponential law was fitted in the range $1\leq -S^*_{ex}\leq 3$ including all isobars. It is worth mentioning that with this fitting range, the exponential law still holds even below the melting line for low pressures roughly $P^*\leq 0.5$.

\begin{figure}[t!]
	\centering
	\includegraphics[scale=0.7]{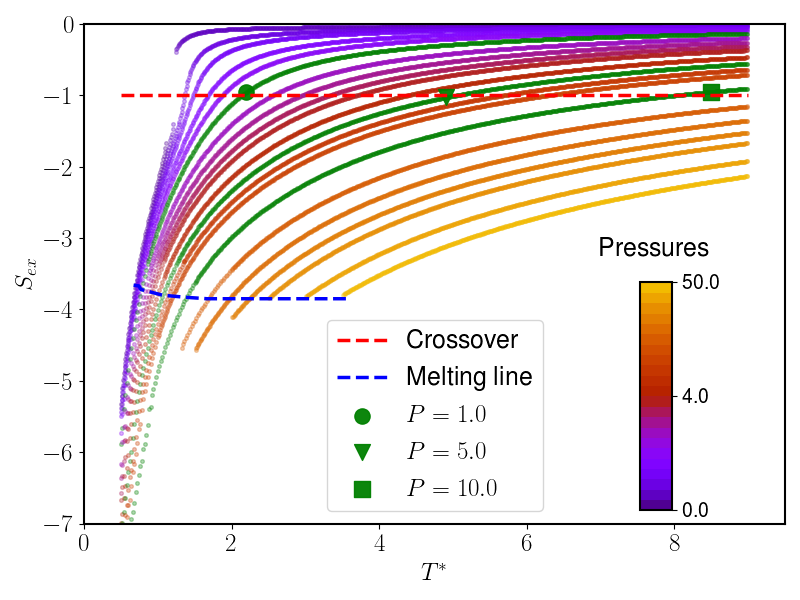}
	\caption{Excess entropy as a function of temperature along isobars for all pressures. The blue dashed line corresponds to the melting line inferred from Ref. \cite{Mastny2007} and the red dashed one marks the crossover at $-S^*_{ex}=1$ (see text). Green symbols represent the onset of free diffusion regime in the potential energy landscape for $P^*=1.0$, $5.0$ and $10.0$.}
	\label{fig:Fig2}
\end{figure}

Fig.\ref{fig:Fig2} displays the temperature evolution of the excess entropy for all considered isobars. The blue dashed line shows the values of $S^*_{ex}$ found for the densities and temperatures corresponding to the melting line taken from Mastny \textit{et al.} \cite{Mastny2007}. Interestingly, $S^*_{ex}$ remains quite constant with a value of $\sim 3.85$ up to $P^*=50$ that highlights once more the importance of excess entropy in describing the phases of matter \cite{Jakse2003}. The dashed red line corresponds to the crossover between the power and exponential scaling laws as described above. Three green curves were chosen to intersect the crossover over the complete range of temperatures investigated, avoiding them being too close to liquid-gas coexistence line and critical point. They correspond to isobars $P^*=1.0$, $P^*=5.0$, and $P^*=10.0$ (see also Figs. S4 S5 and S6 in the supplementary information\cite{Supplement}) that were investigated further by MD to reveal the properties of their underling PEL.
\begin{figure}[t!]
	\centering
	\includegraphics[scale=0.7]{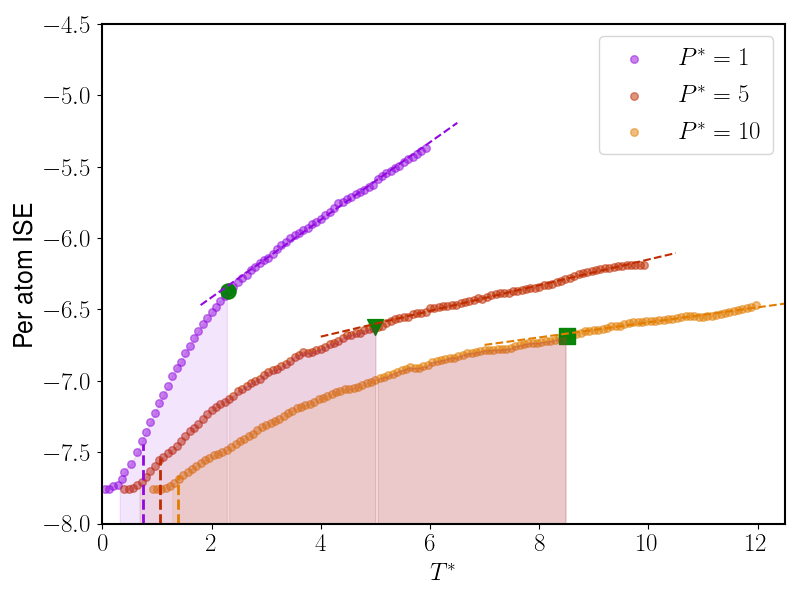}
	\caption{Inherent Structure Energy for $P^*=1.0$, $5.0$ and $10.0$. The filled areas under the curves correspond to the landscape influenced regime with a lower limit corresponding to the dynamical glass transition of $0.32$, $0.67$, and $1.20$, and a upper limit of $2.32$, $5.04$ and $8.52$ respectively for $P^*=1.0$, $5.0$ and $10.0$. The respective vertical dashed lines mark the melting points, namely $0.74$, $1.061$, and $1.379$, after Ref. \cite{Mastny2007}.}
	\label{fig:Fig3}
\end{figure}

For each temperature above the melting line, an average Inherent-Structure Energy (ISE) \cite{sastry1998} over $50$ independent configurations was determined by means of conjugate gradient minimization, bringing the system local minimum of the PEL. Below the melting line, the system was first rapidly quenched with a cooling rate of $0.02$ before calculating the ISEs in the same manner. The corresponding ISE curves for the three pressures are drawn in Fig. \ref{fig:Fig3} for which an additional Stavitky and Golay smoothing  with $10$ points was applied. A typical behavior as a function of temperature \cite{sastry1998} is seen for all the curves, with an almost linear regime at high temperature corresponding to the so-called free diffusion regime that turns to a more pronounced decrease upon further cooling as the system enters the landscape-influenced regime (highlighted by a colored area under the curves). Finally, when the dynamical glass transition is reached, ISE curves remain essentially constant at low temperatures, marking the glassy regime. The turning points between the free diffusion and landscape influenced regime, marked by symbols were determined by a linear regression by least square minimization of the free diffusion regime. Decreasing progressively the hypothetical turning point temperature (\textit{i.e.} increasing the fitting range) the final value was identified when the Pearson correlation coefficient was below a threshold of $0.98$. The corresponding $p$-value of the $t$-statistics with a null hypothesis of a non-linear regression was always below $0.05$. The ISE are seen to decrease with pressure while the range of landscape influenced regime widens.   
  
The excess entropy corresponding to the temperature separating the landscape-influenced and the free diffusion regimes determined above for each of the three pressures are also drawn in Fig. \ref{fig:Fig2} with the symbols. An excellent match is found with the crossover in the excess entropy scaling laws, demonstrating a strong correlation between the two in a wide range of pressures and temperatures. The free diffusion regime is representative of the dilute gas with low density and high temperature fluid  where the PEL is only weakly felt by the system. The latter can be considered as composed of colliding hard sphere particles having a diffusion proportional to the mean free path and thermal velocity. This might be well undertaken in the framework of Enskog's theory \cite{Hansen2006} from which emerges the power law with exponent $2/3$ \cite{Rosenfeld1999}. For $-S^*_{ex} \geq 1$, our results show that Rosenfeld's exponential law might be related to the landscape influenced regime whatever the pressure. The progressive transition region between weak and strong coupling behaviors of the entropy scaling curve around $-S^*_{ex}\simeq 1$ mirrors the one of the ISE curves between diffusion and landscape influenced regimes shown in Fig. \ref{fig:Fig3}. Although not justified in details by Rosenfeld initially,  his macroscopically reduced units involves a general length scale $\rho^{-1/3}$ generally associated to an average distance between particles. It can be interpreted as a mean free path between collisions at low densities that progressively tends to the average inter-particle distance as the density increases. Along this line, a modified excess entropy scale was proposed very recently, unifying the two regimes \cite{Bell2019}.  

While the Landscape influenced regime of the PEL extends to the glass transition by definition \cite{sastry1998}, the exponential seems to hold generally only above the melting point. Fig. \ref{fig:Fig1} indicates however that it extends significantly below the melting line for low pressures, consistently with the PEL view. For higher pressure, probably the high density and low pressure thermodynamic states are less well described by the underlying equation of states \cite{Johnson1993} considered, preventing accurate estimation of the entropy with the ML approach, accuracy known to be crucial \cite{Jakse2016}. Interestingly, MD simulations with the Kob-Andersen binary LJ model \cite{Agarwal2011} show that excess-entropy scaling still holds down to the mode coupling critical temperature, but the exponential law seems to be limited to moderate undercooling only, recalling though that the configuration entropy was used for the scaling in that case. This prompts us to further investigations below the melting line \cite{Bell2020}. 

In summary, a significant body of self diffusion coefficients and potential energies obtained by large scale molecular dynamics simulation was combined to machine learning to efficiently and accurately determine the entropy for temperatures and pressures ranging respectively to  $0.5 < T^* \leq 6.0$ and  $0 < P^* \leq 50.0$ representative of the essential part of the LJ phase diagram for the fluid phase. Our results reveal that excess-entropy scaling discovered by Rosenfeld is a consequence of the properties of the underlying potential energy landscape. This link reflects the complexity of the PEL with local minima fixing the number of accessible states which corresponds to the very definition of the entropy as well as barrier heights fixing the transition rates. Such an intimate relation between the PEL and the dynamics was also demonstrated recently to be quite general for pure liquid metals \cite{Demmel2021}.  Interestingly, as revealed here, the crossover line between the free diffusion and the landscape influenced regimes as well as the melting line within the pressure range investigated here seem to be excess entropy invariant. For dense liquids, it was shown that isomorph invariance might be the appropriate theoretical framework \cite{Dyre2018}. Given the link revealed here, it could be more general as the PEL view offers a generalization ground and opens the way for further theoretical developments and understanding.
	
\section*{Acknowledgments}

One of us, NJ, would like to express deep thanks to Jérôme Schwindling for initial discussions and advices on neural networks. We acknowledge the CINES and IDRIS under Project No. INP2227/72914, as well as CIMENT/GRICAD for computational resources. This work was performed within the framework of the Centre of Excellence of Multifunctional Architectured Materials “CEMAM” ANR-10-LABX-44-01 funded by the “Investments for the Future” Program. Thiswork has been partially supported by MIAI@Grenoble Alpes (ANR-19-P3IA-0003). Fruitful discussions within the French collaborative network in high-temperature thermodynamics GDR CNRS 3584 (TherMatHT) are also acknowledged.

\section*{Supplementary information}
\section{Machine learning: Artificial Neural  Network}
\begin{figure}[t!]
	\centering
	
	\begin{tikzpicture}[thick,scale=1.1, every node/.style={scale=0.9}]
		\node [draw, shape=circle,fill=Blue!30] (x10) at (0,1.2) {$x_1^0$};
		\node [draw, shape=circle,fill=Blue!30] (x20) at (0,-1.2) {$x_2^0$};
		
		\node [draw, shape=circle,fill=Green!30] (y11) at (3,2) {$y_1^1$};
		\node [draw, shape=circle,fill=Green!30] (y21) at (3,1) {$y_2^1$};
		\node (y31) at (3,0) {$\vdots$};
		\node [draw, shape=circle,fill=Green!30] (y41) at (3,-1) {$y_{12}^1$};
		\node [draw, shape=circle,fill=Green!30] (y51) at (3,-2) {$y_{13}^1$};
		
		\node [draw, shape=circle,fill=Yellow!30] (y12) at (6,1.5) {$y_1^2$};
		\node (y22) at (6,0.5) {$\vdots$};
		\node (y32) at (6,-0.5) {$\vdots$};
		\node [draw, shape=circle,fill=Yellow!30] (y42) at (6,-1.5) {$y_{12}^2$};
		
		\node [draw, shape=circle,fill=Red!30] (y13) at (9,0) {$y_1^3$};
		
		\draw[->] (x10) -- (y11);
		\draw[->] (x10) -- (y21);
		\draw[->] (x10) -- (y31);
		\draw[->] (x10) -- (y41);
		\draw[->] (x10) -- (y51);
		
		\draw[->] (x20) -- (y11);
		\draw[->] (x20) -- (y21);
		\draw[->] (x20) -- (y31);
		\draw[->] (x20) -- (y41);
		\draw[->] (x20) -- (y51);
		
		\draw[->] (y11) -- (y12);
		\draw[->] (y11) -- (y22);
		\draw[->] (y11) -- (y32);
		\draw[->] (y11) -- (y42);
		
		\draw[->] (y21) -- (y12);
		\draw[->] (y21) -- (y22);
		\draw[->] (y21) -- (y32);
		\draw[->] (y21) -- (y42);
		
		\draw[->] (y21) -- (y12);
		\draw[->] (y31) -- (y22);
		\draw[->] (y31) -- (y32);
		\draw[->] (y31) -- (y42);
		
		\draw[->] (y41) -- (y12);
		\draw[->] (y41) -- (y22);
		\draw[->] (y41) -- (y32);
		\draw[->] (y41) -- (y42);
		
		\draw[->] (y51) -- (y12);
		\draw[->] (y51) -- (y22);
		\draw[->] (y51) -- (y32);
		\draw[->] (y51) -- (y42);
		
		\draw[->] (y12) -- (y13);
		\draw[->] (y22) -- (y13);
		\draw[->] (y32) -- (y13);
		\draw[->] (y42) -- (y13);
		
		\node at (1.5,2) {$w_{1,1}^1$};
		\node at (1.5,1.35) {$w_{2,1}^1$};
		\node at (1.5,0.95) {$\vdots$};
		\node at (1.5,-0.75) {$\vdots$};
		\node at (1.5,-1.35) {$w_{12,2}^1$};
		\node at (1.5,-2) {$w_{13,2}^1$};
		
		\node (b11) at (3,2.7) {$b_1^1$};
		\node (b51) at (3,-2.7) {$b_{13}^1$};
		
		\node at (4.5,2.2) {$w_{1,1}^2$};
		\node at (4.5,1.5) {$w_{2,1}^2$};
		\node at (4.5,1.05) {$\vdots$};
		\node at (4.5,-0.93) {$\vdots$};
		\node at (4.5,-1.5) {$w_{12,12}^2$};
		\node at (4.5,-2.2) {$w_{12,13}^2$};

		\node (b12) at (6,2.2) {$b_1^2$};
		\node (b42) at (6,-2.2) {$b_{12}^2$};
		
		\node at (7.3,1.15) {$w_{1,1}^3$};
		\node at (7.3,-1.15) {$w_{1,12}^2$};

		\node (b12) at (9,0.7) {$b_1^3$};

	\end{tikzpicture}
	\caption{Schematic representation of the feed-forward neural network built as a densely connected multi-layer perceptrons. The input layer $\{x^0_1;x^0_2\}$ will be fed with density $\rho^*$ and temperature $T^*$ couples of the training set, and the output $y^3_1$ takes the value of corresponding Helmholtz free energy $A^*_{ex}$. The neural network contains two hidden layers with superscript $1$ and $2$ respectively. The first layer is composed of $13$ neurons and the second one $12$. The weights $w^l_{ijk}$ and bias $b^l_i$ are optimized during the training (see text).}
	\label{fig:FigS1}	
\end{figure}
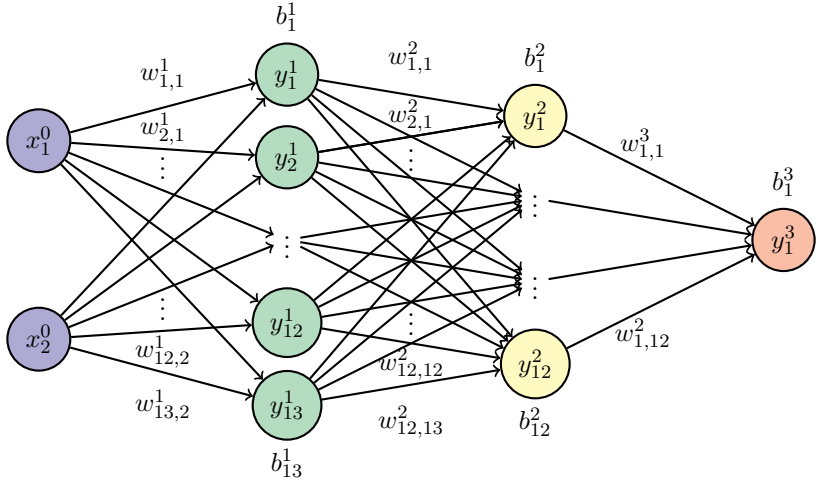

Artificial Neural Networks (ANN) are defined by a network topology which specifies the number of neurons formally named here $y^{l}_{ij}$ and their connectivity through the weights, $w^l_i$. Fig. \ref{fig:FigS1} represents schematically the typical architecture used in the present work, called a Multi-Layer Perceptron (MLP). The weights associated with each node pairs are optimized during the learning process by a Feed-Forward technique, in which each of the $N$ layer within the neural network consists of sets of nodes which receive multiple inputs from the previous layer and pass outputs to the next layer. Here a fully connected network, where every output within a layer is an input for every neuron in the next layer. The corresponding mathematical description is the following: the inputs signals are linearly combined before being activated by function $f$ to give each  output $y^l_i$ of a given fully connected layer $l$ as 
\begin{equation}
	y_i^{l} = f \left(\sum_{j=1}^{N_{l-1}}w_{ij}^l y_j^{l-1}+b_i^l\right), \label{layer_eq}
\end{equation}
where $N_l$ refers to the size of the $l$-th layer, \textit{i.e} the number of neurons in the layer. Note that positive weights enhanced connections while negative weights tend to inhibit the connections. Most of the activation function are choose to have a range in either $[0,1]$ or $[-1,1]$ and modulates the amplitude of the output. The activation function $f$ is applied element-wise and is taken as the hyperbolic tangent form $f(x) = \tanh(x)$. Back-propagation is used to update the network weights and their the gradients.

In a first stage, a training is carried out to find the optimal set of weights and biases that best represents existing data set for the sake of making accurate predictions about new observations. The data set is composed of $16448$ output values of $A^*_{ex}=A_{ex}/\epsilon$ for regularly spaced input couples $(\rho^*,T^*)$ in the phase diagram domains (i) $0.5 < T^* \leq 6.0$ using the Johnson, Zollweg and Gubbins \cite{Johnson1993} equation of states, and $6.0 < T^*\leq 9.0$ using the Thol \textit{et al.} \cite{Thol2016} one. Densities from $0.005$ to $\rho^*=1.2$ were considered thus covering a range of pressures up to $P^*=50$. The typical temperature and density meshes are respectively $0.065\epsilon/k_B$ and $0.015\sigma^{-3}$ The data points in the unstable liquid-gaz region were removed. The complete data set is primarily randomized and scaled with
\begin{equation}
	\texttt{SS}(x) = \frac{x - m(x)}{\sigma(x)}, 
	\label{standard_scaler}
\end{equation}
where $m$ refers to the mean and $\sigma$ to the standard deviation, while $x$ is an input series of the dataset. The training set consists in taking 80\% of the data set, the remaining part is left for the test set. The training set is further split to get a validation set, which amount to 20\% of the training observations. For a given architecture, the optimization of weights and biases is performed using the training data alone, terminating when the validation error begins to increase. Simultaneously, a $L_2$ norm regularization is perform to check for consistency. An early stopping criterion is used on training and validation sets to avoid over-fitting the training data.

\begin{figure}[tb!]
	\centering
	\includegraphics[scale=0.65]{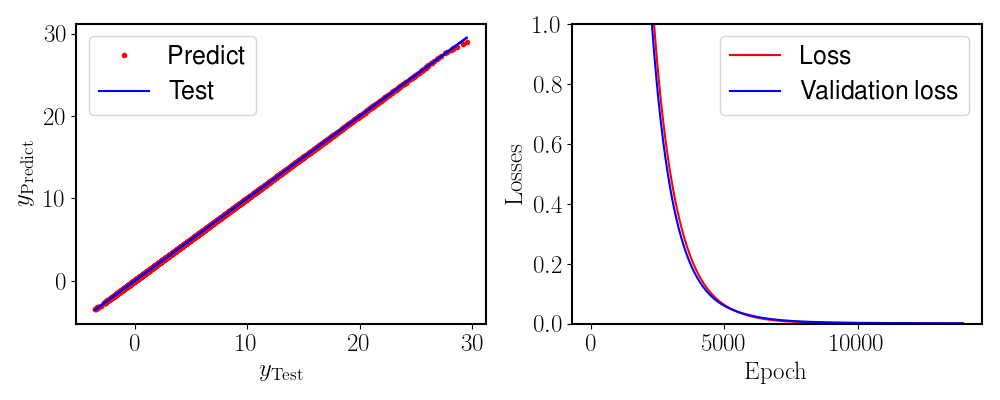}
	\caption{Left panel: Train-test curve showing the quality of the prediction on the test set for the optimized NN with 13 and 12 neurons respectively in the first and the second hidden layers. The blue line represents the known output of the Helmholtz free energies and the red dots the values predicted against the known ones. Right panel: Evolution of the MSE losses as a function of number of epochs for the training and validation sets.}
	\label{fig:FigS2}
\end{figure}

\newpage
This training stage is repeated with various ANN architectures to find the optimal one that can capture numerically at best the functional dependence of the data. The minimization of the mean-square error (MSE) is carried out with stochastic gradient descent using the Adam optimization algorithm \cite{Hastie2009} giving a measure of the loss with a constant learning rate of $0.001$, $\beta_1=0.9$, $\beta_2=0.999$ and $\varepsilon=10^{-8}$. The early-stopping was performed with maximum loss variation of $10^{-6}$ and a patience of $45$ epochs. The least MSE loss is obtained for an architecture of $13$ and $12$ neurons in the first and second hidden layers respectively as depicted in Fig. \ref{fig:FigS1}. A cross-validation is performed over $10$ independent trainings gives a loss of $(2.3\pm0.4)\times 10^{-4}\epsilon$ on the per atom Helmholtz free energy. The left panel of Fig. \ref{fig:FigS2} displays the predictive ability of the model on the unseen data of the test set. The quality is very high on the whole range of the free energy with a small deviation at the very end of the large values for which there is less data. The right panel of Fig. \ref{fig:FigS2} shows the evolution of the MSE loss as a function of epochs. The typical duration of the training period was about $10000$ to $15000$ epochs.

\section{Test of the Neural Network Model}

\begin{figure}[tb]
	\centering
	\includegraphics[scale=0.5]{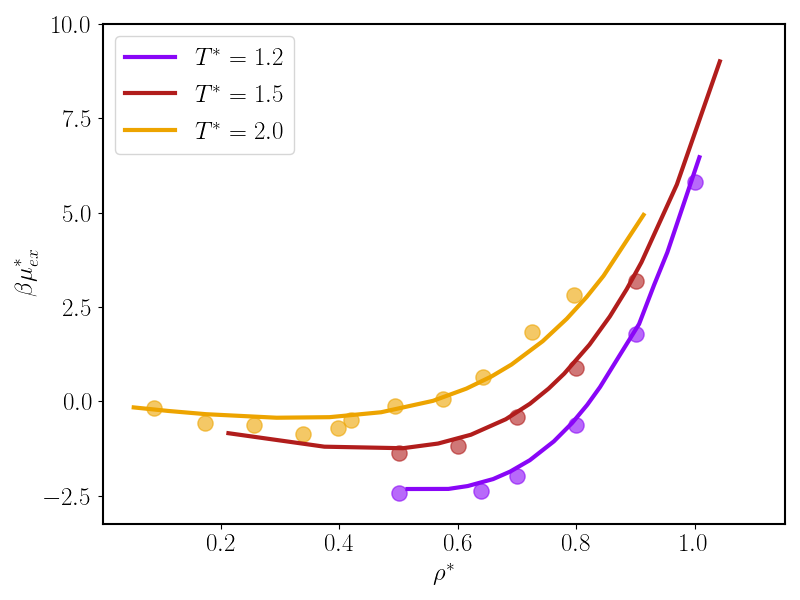}
	\caption{Excess chemical chemical potential reduced by the temperature ($\beta=1/T^*$ ) as a function of density. Predictions of the NN (solid lines) are compared to the direct molecular dynamics simulations (filled circles) of Ref. \cite{Rowley1994} for isotherms $T^*=1.5$ and $2$, and Ref. \cite{Lee1996} for $T^*=1.2$.}
	\label{fig:FigS3}
\end{figure}

From the predicted excess Helmholtz free energy $A^*_{ex}$  the excess entropy can be determined readily from MD simulations by the thermodynamic relation 
\begin{equation}
	S^*_{ex} =  {U^*_{ex} - A^*_{ex}\over T^*}
	\label{Eq:sexT}
\end{equation}
in which $U^*_{ex}$ is the excess energy. In order to assess the reliability of the approach, results of the excess entropy is compared to independent published MD dynamics simulation\cite{Rowley1994,Lee1996} through the  excess chemical potential $\mu_{ex}$ using the standard relation \cite{Jakse2003}
\begin{equation}
	S_{ex}=\frac{\beta P}{\rho }-1+\frac{\beta U_{ex}}{%
		\left\langle N\right\rangle }-\beta \mu_{ex}.  
	\label{sexT}
\end{equation}
where $P$ is the pressure. Fig. \ref{fig:FigS3} shows the excess chemical potential for several isotherms. A good agreement is found for all of them given the fact that the small number of atoms in those simulations tends to underestimate the chemical potential \cite{Rowley1994}

\section{MD simulation results along isobars}

\begin{figure}[tb]
	\centering
	\includegraphics[scale=0.8]{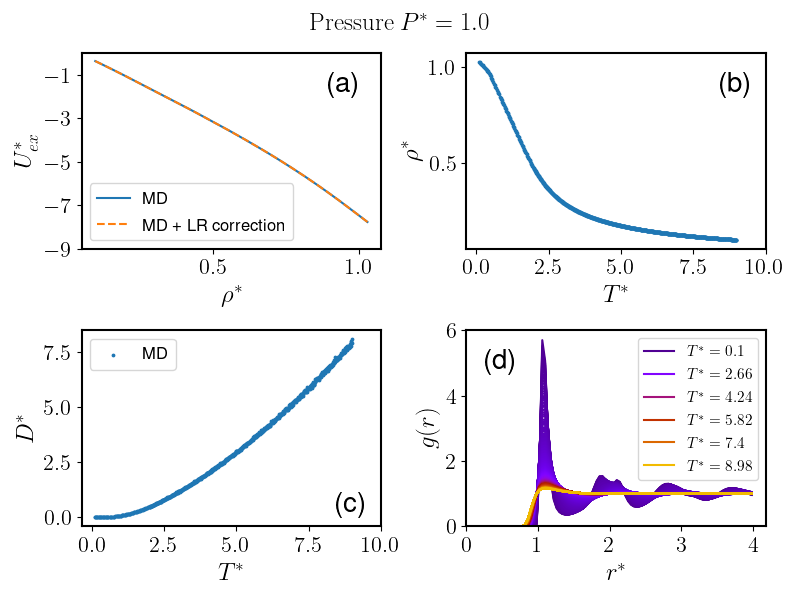}
	\caption{Molecular dynamics simulation results along isobar $P^*=1$, namely excess potential energy (a), density (b), self-diffusion coefficients (c) and pair-correlation functions (d).}
	\label{fig:FigS4}
\end{figure}

\begin{figure}[tb!]
	\centering
	\includegraphics[scale=0.8]{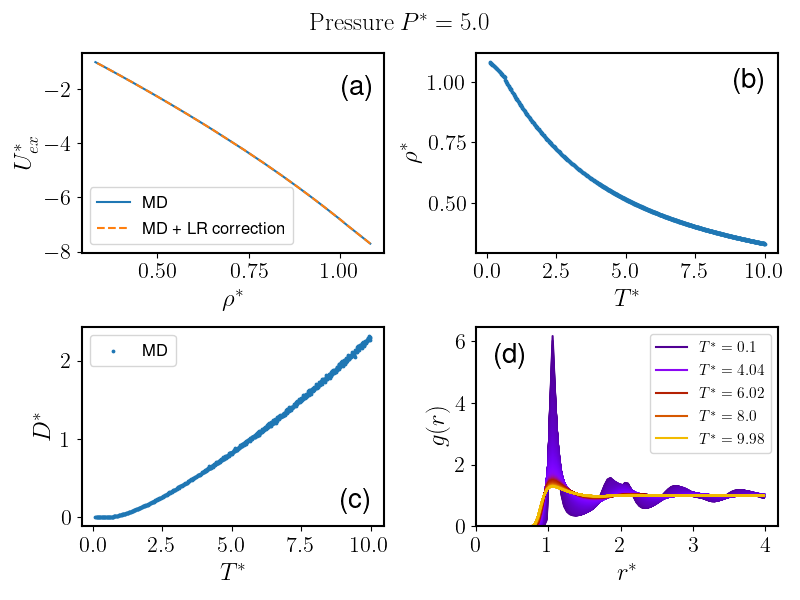}
	\caption{Same as in Fig. \ref{fig:FigS4} for isobar $P^*=5$.}
	\label{fig:FigS5}
\end{figure}

This Section presents the results along isobar $P^*=1.0$, $5.0$, and $10.0$ respectively in Figs. \ref{fig:FigS4}, \ref{fig:FigS5}, and \ref{fig:FigS4}. Various quantities were collected every temperature spaced by temperature step $\Delta T=0.02$. Below the melting point, the system was cooled down with a fast quenching rate of $Q=0.02$ to avoid spontaneous crystallization. The dynamical glass transition occurs at the crossover seen on the temperature evolution of the density (panels b) and the Diffusion (panels c). in panels (a) the excess potential energy with and without long range corrections are compared, showing that they are always very small. For each generated state point, Inherent Structure Energies\cite{stillinger1995,sastry1998} were determined from a conjugate gradient minimization on $5$ independent configurations. The needed data from these isobars as well as all others were stored in a data set to determine in a post treatment the excess entropy with the machine learning approach.            

\begin{figure}[tb!]
	\centering
	\includegraphics[scale=0.8]{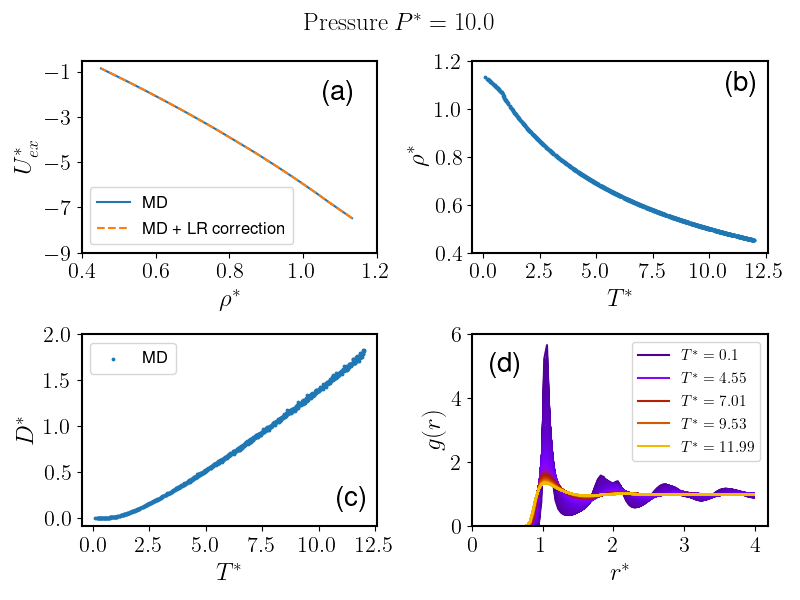}
	\caption{Same as in Fig. \ref{fig:FigS4} for isobar $P^*=10$.}
	\label{fig:FigS6}
\end{figure}

\end{document}